# THE STOCK MARKET AS A GAME:
# AN AGENT BASED APPROACH TO TRADING IN STOCKS

Eric Engle

"If ever there were a field in which machine intelligence seemed destined to replace human brainpower, the stock market would have to be it. Investing is the ultimate numbers game, after all, and when it comes to crunching numbers, silicon beats gray matter every time. Nevertheless, the world has yet to see anything like a Wall Street version of Deep Blue, the artificially intelligent machine that defeated chess grand master Gary Kasparov in 1997. Far from it, in fact: When artificial-intelligence-enhanced investment funds made their debut a decade or so ago, they generated plenty of media fanfare but only uneven results. Today those early adopters of AI, like Fidelity Investments and Batterymarch Financial, refuse to even talk about the technology...Data flows in not just from standard databases but from everywhere: CNN, hallway conversations, trips to the drugstore. ‚Unless you can put an emotional value on certain events and actions, you can't get the job done.' Naturally, investors don't process this hodgepodge of inputs according to some set of explicit, easily transcribed rules. Instead, the mind matches the jumble against other jumbles stored in memory and looks for patterns, usually quite unconsciously. ‚Often, great investors can't articulate the nature of their talent. They're like pool players who make incredible trick shots on intuition.' Fine for them, but how do you code that?"[1]

A Taxonomy of Games

A game can be defined as a set of rules (conditionals) with one or more goals (also conditionals) with an outcome of „win" or „loss" depending on whether the conditionals are fulfilled.[2] Games can either be positive sum, zero sum, or negative sum.[3] Positive sum games, such as trading goods, are games in which all parties to the game are, in absolute terms, better off as a result.

---

[1] Carla Fried, „Can technology build a better Buffett?", (February 2004)
http://www.cnn.com/2004/TECH/ptech/02/12/bus2.feat.buffett.ai/index.html
[2] Wikipedia, „Game", (2004)
http://en.wikipedia.org/wiki/Game
[3] Wikipedia, „Non-Zero Sum", (2004)
http://en.wikipedia.org/wiki/Non-zero-sum

Trading of goods is generally a positive sum game: each party has a good the other good use that it cannot. Both parties are both better off because of the trade. Negative sum games are games in which all parties, in absolute terms, are worse off.[4] War is an example of a negative sum game. All participants in a war suffer dead and maimed persons and waste riches in mutual destruction. War is often erroneously represented as a zero sum game. In a zero sum game any improvement of one participant's position results in a deterioration of the other participants position.

Just as war is sometimes fallaciously represented as a zero sum game – when in fact war is a negative sum game – stock market trading, a positive sum game over time, is often erroneously represented as a zero sum game. This is called the „zero sum fallacy"[5] – the erroneous belief that one trader in a stock market exchange can only improve their position provided some other trader's position deteriorates.[6] However, a positive sum game in absolute terms can be recast as a zero sum game in relative terms. Similarly it appears that negative sum games in absolute terms have been recast as zero sum games in relative terms: otherwise, why would zero sum games be used to represent situations of war? Such recasting may have heuristic or pedagogic interest but recasting must be clearly explicited or risks generating confusion.

Availability of Information

Games can also be classified according to how much information is available to players. In a game with perfect information all states are known to all players at all times. Chess or Go are examples of games with perfect information. In a game with imperfect information in contrast, at least some information is not known to some (possibly all) of the players at least some of the time. Card games generally are examples of games with imperfect information.[7] Information may be further distinguished into private knowledge (information known only to one player); public knowledge (information known to all players; share information (known by two or more

---

[4] Brad Spangler, „Positive-Sum, Zero-Sum, and Negative-Sum Situations" (2003) http://www.intractableconflict.org/m/sum.jsp
[5] Wikipedia, „Zero-Sum Fallacy", (2004) http://en.wikipedia.org/wiki/Zero-sum_fallacy
[6] Wikipedia, „Non-Zero Sum", (2004) http://en.wikipedia.org/wiki/Non-zero-sum
[7] Brad Spangler, „Positive-Sum, Zero-Sum, and Negative-Sum Situations" (2003) http://www.intractableconflict.org/m/sum.jsp

players); completely unknown by any player.[8]

Determinicity

Games can also be classified depending on whether they are subject to random influences. Deterministic games, such as chess or go, have no random elements. Most card games in contrast have random aspects. Interestingly, games with random factors generally also include imperfect information, and deterministic games usually have perfect information. However examples of deterministic games with imperfect information such as Stratego can be found. Similarly games with perfect information and random elements such as backgammon also exist.

Solved Games[9]

Games can also be described as „solved" or „unsolved". A game can be solved in at least three senses:

In the weakest sense („ultra-week") a game is solved if, given an initial position and perfect play on both sides we can predict whether the first player to move will win, lose or draw.

A more usual meaning of „solved game" is to define the game as solved where an algorithm exists which will secure a win or draw for a player from the initial position regardless of any move by an opponent. This is the „weak" definition of a solved game.

The „strong" definition of a solved game is defined as having an algorithm which can produce the best play possible from any position at any time within the game. Thus even in mid game, even after mistakes have been made by either side the algorithm still returns the perfect play. It is always possible, but often computationally intractable, to produce such an algorithm in

---

[8] Wikipedia, „Zero-Sum Fallacy", (2004)
http://en.wikipedia.org/wiki/Zero-sum_fallacy
Brian Graney, How Money Is Made in the Market", (2000)
http://www.fool.com/news/foth/2000/foth000912.htm

[9] Wikipedia, „Solved Board Games", (2004)
http://en.wikipedia.org/wiki/Solved_board_games

games with a finite number of positions.[10]

Symmetry

Games can also be described as symmetric or asymetric. Symetric games are those where the players have equal resources and where each of their moves effectively „mirrors" those of the opponents.
 In a symmetric game, a move that is good for white is bad for black and vice verse. In contrast, in asymmetric competitions the resources of the parties are unequal.[11]

Dominant Strategy

Dominant strategies emerge in a game where a party has a move that always leads to a winning position regardless of the moves undertaken by their opponent.[12]
 Prisoner's dilemma is an example of a game where each player has a dominant strategy, namely to implicate their co-conspirator.[13]

Stock Market Games

In stock market games the objective of each player is to maximize their wealth. Wealth maximization as a goal can be undertaken either cooperatively or conflictually. However the use of war or theft to maximize individual wealth not only reduces overall social wealth it also is ultimately ineffective since it destroys any incentive to productivity. Cooperative strategies of wealth maximization are much more effective: each party gives up some of their surplus to

---

obtain that which they do not have but need or at least want. Further, cooperative strategies encourage investment in the future because expectations of stability are created. Finally, cooperative strategies encourage specialization of labor and ultimately introduce economies of scale.

However, while economic games are in absolute terms clearly positive sum[14]
 – and this has been scientifically proven by Adam Smith[15]
 and Ricardo[16]
 - we can recast them as zero sum games in relative terms. This reintroduces the sense of competition making the game more interesting for all participants.

The goal then of a stock market game is to not merely maximize wealth but rather to maximize wealth faster than one's competitors. What are the properties of a stock market?

In the stock market we are presented with nearly perfect information. We could know the trading history of all stocks. We even know the trading patterns of „insider" traders who are subject to disclosure requirements when they trade. However the problem is not getting the information – rather the problem is there is too much information! A major problem involved in modelling the stock market is gathering data and putting it into a useful knowledge base. [17]
While we know past information nearly perfectly we do not know the intentions or opinions of our opponents. We do not know what the portfolio of our opponent looks like. The information is nearly perfect but it is also, aside from inside traders, anonymous.

Is the stock market deterministic or random? In fact the stock market is deterministic. Prices rise and fall based on the laws of supply and demand. However, again, the vast amount of information influencing the economy makes modeling the stock market as a whole difficult. The price of oil, inflation, interest rates, unemployment rates, wars, strikes, new inventions, rates of taxation, trade agreements -  all influence the stock market sometimes obviously, sometimes subtly. For example, a stock market will appear to perform well during inflation - but in reality the growth is merely a reflection of the devaluation of the currency! This is the current case of the U.S. stock market. The inflation of the dollar is making the stock market there look more profitable than it is.

---

[14] Marc. T. Law and Fazil Mihlar, "Debunking the Myths: A Review of the Canada-US Free Trade Agreement and the North American Free Trade Agreement" 11 Public Policy Sources (2000), http://oldfraser.lexi.net/publications/pps/11/conclusion.html ; Bipul Chatterjee, "Trade in Services - Cul de Sac or the Road Ahead!" CUTS Briefing Paper, July 1997, Number 7, http://www.cuts-international.org/1997-7.htm

[15] Adam Smith, An Inquiry into the Nature and Causes of the Wealth of Nations, (1776) Chapter 2 http://www.bibliomania.com/2/1/65/112/7058/1/frameset.html; "Of Restraints upon the Importation from Foreign Countries of such Goods as can be produced at Home" http://www.bibliomania.com/2/1/65/112/7058/1/frameset.html

[16] David Ricardo On The Principles of Political Economy and Taxation London: John Murray, Albemarle-Street, 1817 (third edition 1821) Chapter 7 "On Foreign Trade" http://phare.univ-paris1.fr/textes/Ricardo/Principles/prin1.txt

[17] See, e.g., TDM Research „Our Models" http://www.tdmresearch.com/OurModels.htm

Stock market trading can also be said to be asymmetric. Some players are very very rich, others are not rich at all. Some have access to information, others even if they have access to information do not know how to use it.

Stock market trading is for this reason an unsolved and likely unsolvable game. The information, theoretically perfect, is practically intractable. Further the number of possible moves (purchase and sale of given securities) is infinite.

Interest of Artificial Intelligence in Stock Market Trading
Existing Stock Market Games

Stock market games exist both online, [18] [19]

and offline[20]

including open source projects.[21]

The objective of stock market games is for players to learn about investment strategies safely. According to Chris Crawford the fact that games allow us to safely experiment with models of

---

[18]    SmartStocks.com „Stock Market Simulation Game" (2004)
        http://www.smartstocks.com/login.html
[19]    MyStocks, „Global Stock Market Game" (2004)
        http://investsmart.coe.uga.edu/C001759/stocksquest/mystocks.htm
[20]    Computer-Game.us, „Stock Market Game", (2004)
        http://www.computer-game.us/strategy_war/wall_street_raider.htm
[21]    Source Forge, Open Source Artificial Stock Market (2004)
        http://sourceforge.net/projects/artstkmkt/

reality explains the pedagogic utility of games.[22]

These games are of commercial interest – for example, the German Postbank currently uses a stock market game to attract clients.[23]

Automated Trading

Artificial intelligence algorithms for stock trading are not only of academic or ludic interest. They are of real importance in actual stock market trading. Automated stock trading is a part of daily stock trading today.[24]

Investment companies develop and deploy automated trading strategies.

Neural Networks[25]

A neural network is a cognitive model of a brain which can be trained through trial and error to achieve a certain state. Interestingly, most AI modelling of stock markets at present is not using reinforcement learning or opponent modelling. Rather neural networks seem to be the centre of current research and writing on artificial intelligence in the stock market.[26]

Neural networks have commercial application in stock market trading[27] where there are numerous programs available for end users to predict stock market performance.

Artificial Intelligence Methods which can be applied to Stock Trading

---

[22] Chris Crawford, The Art of Interactive Design, From Concept to Reality. No Starch Press (2002).

[23] PostBank „Easy Trade" (2004)
http://www2.easytrade.de/

[24] Sungard, „Products and Services (2004)
http://www.sungard.com/products_and_services/sts/brass/solutions/automatedtrading.htm

[25] For a list of numerous articles on neural networks as tools for trading stocks see: Bo Qian „Research", http://www.arches.uga.edu/~qianbo/Research.htm

[26] For an example of the commercial application of neural networks to stock market trading see: AAStocks.com, „Artificial Intelligence Applied to Stock Trading"
http://www.aastocks.com/eng/Education/ai_stock_trading.asp

[27] See, e.g. „Artificial neural network software for stock market & trading forecasts. Market forecasting Software,
http://www.ozgrid.com/Services/neural-network-software.htm

Minimax? Alpha Beta? Expectimax?

The minimax algorithm holds that we should take those moves which maximize our wins and that we should presume that our opponent will take those moves which minimize his losses. In a zero sum game where the moves can be represented using a tree structure minimax is very useful. But the moves in a stock market are simply sales and purchases of stock. Moreover we are trying to anticpate the movement of the market as a whole and the movement of a particular stock. Thus minimax may not be applicable. This is all the more true because economic exchanges are usually positive sum: a move which maximizes my gains and minimizes my losses will not necessarily minimize your gains and maximize your losses. Since the only movements we are interested in are individual sales or purchases of a stock or estimates of the agregate market we are not looking at searching a tree for right or wrong moves. Rather, we are rather evaluating a stock based on its fundamentals (fundamental analysis) or the market as a whole (technical analysis). Since no tree is being searched we also cannot usefully apply alpha-beta pruning to limit the size of our search space – we are not searching a tree with nodes and leaves. Similarly, while we may wish to use pseudo-random elements to represent the market's fluctuations, since we are not searching a tree a probabalistic approach to minimax – expectimax – is not really useful in stock analysis.

We could of course coerce our representation of the stock market into such a form. For example, we could focus on two traders in the wildly fluctuating futures market with option to put or call trades and to sell long or short. However this is of less interest: only very experienced investors play the futures markets because they are extremely risky. Rather than trying to fit a stock market game to the constraints necessary to a board game we should let our model reflect reality. A much more realistic and useful model, presented by this author, focusses only on the ordinary trading of stocks, not on options or futures and can thus safely ignore put, call, limit, and stop loss orders influence on trading.

Machine Learning

One possible method which we could apply to our stock market sales or buying algorithms would be machine learning. In machine learning we „reward" our algorithm when it sells profitably and „punish" it when it's purchase is unprofitably (or even when it underperforms the market average). [28] Machine learning attempts to develop algorithms which learn to recognize recurring patterns and to improve performance based on experience. [29] Clearly such methods can be applied to algorithms for the purchase or sales of stock, likely looking more to technical analysis (examining the market) than fundamental (examining the statistics of this particular company) analysis.

Reinforcement Learning

Reinforcement Learning is a type of machine learning. It uses feedback (known as the reinforcement signal) to tell the software agent when it has performed as desired. Behaviors can be learned once or continually adapt over time. Proper modelling of problems allows reinforcement learning algorithms to converge to an optimum solution. [30] The reinforcement signal „reflects the success or failure of the entire system after it has performed some sequence of actions. Hence the reinforcement signal does not assign credit or blame to any one action (the temporal credit assignment problem), or to any particular node or system element (the structural credit assignment problem)". [31]

Reinforcement learning should be distinguished from supervised learning where feedback occurs after each action. Supervised learning methods rely on error signals at output nodes and train on a fixed set of known examples – and that is only a partial model for learning. Where there is no external algorithm to provide feedback, the algorithm must somehow modify itself to achieve desired results – using reinforcement learning. [32]

---

[28] ************ATIS Committee T1A1, " Machine Learning" (2004)
http://www.atis.org/tg2k/_machine_learning.html
[29] Computer User „Machine Learning" (2004)
http://www.computeruser.com/resources/dictionary/definition.html?lookup=2949
[30] Alex J. Champandard, "Reinforcement Learning" (2002)
http://reinforcementlearning.ai-depot.com/Intro.html
[31] David Finton, "Reinforcement Learning" (1994)
http://www.cs.wisc.edu/~finton/what-rl.html

Opponent Modeling[33]

Opponent modelling is also very relevant to stock market analysis. It is clear that there are various investment strategies – bears, who are sceptical about market performance, bulls who are enthusiastic about market performance, blue chip investors, who seek steady certain gains, and speculators who are willing to take high risks in the hope of great rewards. Each of these strategies is in fact appropriate to a certain investor. Opponent modelling could be used to tell us how the market will behave – if we know the strategies of our opponents, which is not at all certain.

But even if we do not know what the strategies of individual market participants are we may be able to use oppoenent modelling to help predict how the market moves. Say we know one fourth of all market participants are blue-chip investors, buying only stocks based on their dividends, and we know the remainder of the market is equally divided between three types of investors: bears, bulls, and risk takers. This may be useful to help us to model the movement of the market and to determine whether to buy or sell a given stock at a given price.[34]
Interestingly, opponenet modelling has been shown to be superior to MINIMAX if the opponent modelling algorithm has enough time to develop an accurate model of the opponent![35]

---

[32]    University of Massachussets, Amherst „Glossary of Terminology in Reinforcement Learning" (2004)
        http://www-anw.cs.umass.edu/rlr/terms.html
[33]    H. H. L. M. Donkers, J. W. H. M. Uiterwijk, H. J. van den Herik, „Admissibility in opponent-model search" Information Sciences, Volume 154,  Issue 3-4  (September 2003) http://else.hebis.de/cgi-bin/sciserv.pl?collection=journals&journal=00200255&issue=v154i3-4&article=119_aios&form=pdf&file=file.pdf
[34]    University of Massachussets, Amherst „Glossary of Terminology in Reinforcement Learning" (2004)
        http://www-anw.cs.umass.edu/rlr/terms.html
[35]    Bo Qian „Research",
        http://www.arches.uga.edu/~qianbo/Research.htm

Agents

An agent is „A system that is embedded in an environment, and takes actions to change the state of the environment." [36]
 Agents have sensors to percieve environment states
 and affectors to influence it. States are a representation of the history of a system which in turn determines the evolution of the system.[37]
Agents can be combined with opponent modelling. For example we could create agents as opponents which implement a trading strategy. These agents could even have learning functions to allow them to change their trading strategy based on how they perform compared to the market, other agents or the human player.[38]

In an actor critic architecture one agent would execute trades while another determines whether the trade was a good one[39]
 In addition to the „trading" agents, executing „bearish" or „bullish" strategies a „critic" agent could evaluate the results of other agents to try to determine the optimum trading strategy. This agent could then act as the critic to other agents in an actor-critic architecture.

---

[36] For an agent based approach to market analysis which models the market as a set of agents see: Sérgio Luiz de Medeiros Rivero, Bernd Heinrich Storb, Raul Sidnei Wazlawick, „Economic Theory, Anticipatory Systems and Artificial Adaptative Agents", Brazilian Electronic Journal of Economics Vol. 2 No. 2. Their model has numerous agents. Agregate behavior emerges from individual behavior. The agents antipate the future of the system. Thus the diverse agents are adaptive, autonomous and anticipatory.
    http://www.beje.decon.ufpe.br/rivero/rivero.htm

[37] Shyam Sunder, "A computer simulation model for portfolio strategy formulation", Proceedings of the 10th conference on Winter simulation - Volume 2 (December 1978)
    http://portal.acm.org/ft_gateway.cfm?id=807627&type=pdf&coll=ACM&dl=ACM&CFID=34786322&CFTOKEN=50371462

[38] Id. at p. 945.
    Id. at p. 952.
    Id. at p. 949

[39] Id. at p. 945.
    Id. at p. 952.
    Id. at p. 949

Stock Valuation Strategies

There are roughly speaking three tools for analysing the value of a stock. Technical analysis (TA) looks not at the company, but at the market. [40] Technical analysis evaluates the stock based on its sales prices in the past (opening price, closing price, high, low, trading volume). I think this is a good tool for analysing the value of a stock on a given day – unless exogenous factors such as war or other disaster intervene! The other main tool is fundamental analysis. Fundamental analysis (FA) is much more conservative but also more scientifically well founded. In FA we look at the „hard values" of the company. How much has it sold? Were its sales profitable? What is the net value of the company? How much debt does the company have? What is the ration of the share price of the company to the book price of the company? What is the ratio of the price of the company to the earnings of the company? Fundamental analysis is much more exacting. It requires us to understand whether the company is on solid footing and why. Technical analysis alone cannot reveal when a company is undervalued or overvalued. Fundamental analysis can tell us when a company is undervalued (which we would then buy) or when it is overvalued (in which case we must not buy it, rather we should sell). Fundamental analysis is the basis of the investment strategy of Warren Buffett, one of the world's richest men and the world's best stock market trader.[41]

A third approach, which seems very unwise to me, is the „efficient market hypothesis" (EMH). EMH proposes that because stock market information is almost all publically available that the stock market is in a situation of perfect knowledge. Consequently, according to EMH all information is already contained in the current stock price. There are several problems with this.

---

[40] For a listing of examples of AI in technical analysis – especially neural networks – see, Galateia corporation "Primers and Bibliographies" (2001)
    http://www.voicenet.com/~mitochon/linksource/ai00002.htm

[41] „Intrinsic value is an all-important concept that offers the only logical approach to evaluating the relative attractiveness of investments and businesses." Warren Buffett, "An Owner's Manual" (1996)
    http://www.berkshirehathaway.com/2001ar/ownersmanual.html

While stock market information is largely public it is not able to be digested by any one actor or even any one company. Thus though information is nearly perfect but there is a vast amount of hidden information. Further, information is not perfectly available.: there is plenty of imperfect information out there – false or misleading analysis, undisclosed large trading and insider trading for examples. Information is not instantaneous nor cost free. Finally, EMH does not provide us any algorithm to determine whether to buy or sell a stock. We would never buy or sell a stock if we took EMH seriously because the price of the stock could never be overvalued or undervalued. The fact that investors like Buffett and Soros consistently outperform the market refutes the random walk theory of the EMH.

Existing Literature

This paper is focussing on agent based stock market trading programs. Most contemporary literature uses neural networks to represent the stock market or trading in particular stocks[42] but that is not the focus of this paper.

[43]

Early work on agent based approaches (Sunder, 1978) started from the problem how to balance between risk and return of large portfolios of institutional investors -- portfolio management. Sunder did not look at fundamental analysis because he wanted his work to be more accessible to less experienced investors. Sunder sees the investment goals of an investor as constraints on the investment algorithm. Sunder implies that constraints can help us model our investment decisions using AI. He describes basic portfolio theory (a portfolio can have some risky stocks provided these are counterbalanced by more stable ones), that risk and return are positively associated, and that a portfolio cannot always be expected to always beat the market;[44] Modern portfolio theory seeks to manage and limit risk rather than to maximize returns.[45] Sunder's model considers factors not considered by the model presented by this author: reinvestment of income (dividends) earned from stocks and transaction costs.[46] Reinvestment can be ignored in a stock trading program – unlike a portfolio management program. Indeed, Sunder states that his model is designed for portfolio management and not for analysis of an individual stock.[47] Similarly, electronic stock market trading only costs at most 20 euros per transaction. Even for smaller investors transaction costs can be safely ignored.

---

[42] For a listing of examples of AI in technical analysis – especially neural networks – see, Galateia corporation "Primers and Bibliographies" (2001)
http://www.voicenet.com/~mitochon/linksource/ai00002.htm

[43] Bo Qian „Research",
http://www.arches.uga.edu/~qianbo/Research.htm

Other early work also examined the stock trading from an agent viewpoint. Ying, Bromberg and Solomon note that the prevailing view is that the stock market follows a random walk.[48] This view if true would imply that there is no point in technical analysis or indeed in stock analysis at all since a random trading strategy is just as effective. This seems to be an outgrowth of the efficient market hypothesis (EMH) criticized earlier in this paper. Ying et al. do however correctly note that the stock market correlates to the economy as a whole[49] – and thus our model of the market must include a model of the economy, which is a sensible way to perform technical analysis. Because investors like Buffett and Soros do outperform a random walk – through fundamental analysis - we must also consider the economy as a whole, as Ying *et al.* do.

Ying *et. al.* Present interesting information regarding technical analysis, namely:

A small volume in trading correlates to a price decline

A heavy volume in trading correlates to a rise in price

A large increase in olume correlates to a large change in price (either up or down!)

A large volume of trading on day 1 results in a price increase on day 2

---

[44] Shyam Sunder, "A computer simulation model for portfolio strategy formulation", Proceedings of the 10th conference on Winter simulation - Volume 2 (December 1978)
http://portal.acm.org/ft_gateway.cfm?id=807627&type=pdf&coll=ACM&dl=ACM&CFID=34786322&CFTOKEN=50371462
[45] Id. at p. 945.
[46] Id. at p. 952.
[47] Id. at p. 949
[48] Charles C. Ying, Neil B. Bromberg, Martin K. Solomon, "Toward a simulation model of the stock market" Proceedings of the 5th conference on Winter simulation (January 1971) p. 126.
http://portal.acm.org/ft_gateway.cfm?id=811432&type=pdf
[49] Id. at p. 126.

If volume has been decreasing for 5 consecutive days, price will decline for the following four trading days.

If volume has been increasing for 5 consecutive days, price will rise for the following four trading days.[50]

Moreover, they then prove these empirical observations deductively using an example of four traders! However while their technical analysis seems solid, their fundamental analysis is less exciting and not really worthy of reproducing as it does not consist of a solid anylsis of key values. [51]

These facts are not integrated in the agent model I propose but could be in a future version.

Other early research on agent based stock trading focussed on the irrationality of market participants. Krolak, Berquist, Conn and Gilliland also believe that an effective trading model should be able to learn from its mistakes to improve its investment strategy.[52] Krolak *et al.* also consider the relevance of stop-loss orders on trading, a factor ignored in my model. A stop loss basically says „if the value of this stock declines below X then sell the stock immediately at the best possible price". Automated trading with stop losses via artificial intelligence appears to have been in part responsible for the stock market crash of October, 1987. [53]

Later research in agent based AI also looked at irrationality of models – namely, how a rational agent responds to the errors in another agents model of the world (Agent A is rational – Agent B is not) (Ya'akov Gal, Avi Pfeffer). [54]

Only recently does the research begin to try to develop particular algorithms of stock analysis and purchase. For example, Ronggang Yu and Peter Stone develop what they call „the reverse strategy" – namely, purchasing a stock when its price is falling and selling it when its price is rising. [55] Of course this seems counterintuitive, at least to an inexperienced investor. If the stock is declining than surely it will decline further is the enthymatic presumption. However if we consider the maxim „buy low, sell high" we can understand why their algorithm works. We wish to sell our stock at the highest possible price. That will occur when many other people are buying the stock, thus when the price of the stock is rising. Similarly, we wish to buy the stock at the lowest possible price, namely when everyone else is selling. However their algorithm does not pick out the most advantageous time to buy or sell, namely at troughs and peaks respectively. [56] The reverse strategy, while it does not (and perhaps cannot) pick out the best point (the deepest troughs for buying, the highest peaks for selling) will pick out points at which to buy and sell that, presuming the stock is profitable in the long run will eventually yield profitable sales opportunities. The algorithm may not be optimal but it is profitable. They include a pseudocode for their algorithm, reproduced below:

```
while time permits
lastPrice := getLastPrice();
currentPrice := getCurrentPrice();
if currentPrice > lastPrice then
        placeOrder(Sell, currentPrice, volume);
if currentPrice < lastPrice then
        placeOrder(Buy, currentPrice, volume);
```

---

50     Id. at p. 125.
51     Id. at p. 129.
52     P. Krolak, R. Berquist, R. Conn, H. Gilliland, "A simulation model for evaluating the effectiveness of various stock market strategies" Proceedings of the 6th annual conference on Design Automation (January 1969) p. 352.
53     Raymond Kurzweil, „Machine Intelligence: The First 80 Years", (1991)
    http://www.kurzweilai.net/meme/frame.html?main=/articles/art0245.html?m%3D10
54     Ya'akov Gal, Avi Pfeffer, "A language for modeling agents' decision making processes in games" Proceedings of the second international joint conference on Autonomous agents and multiagent systems (July 2003) p. 265..
    http://portal.acm.org/ft_gateway.cfm?id=860618&type=pdf&coll=GUIDE&dl=GUIDE&CFID=34254303&CFTOKEN=84403697

Information Theory

Another problem that the more recent research has had to deal with is the question of surplus information. The stock market information is freely available and nearly in real time. However, many traders act irrationally or are uninformed. Thus the market is not perfectly rational. Yue,

---

[55] Ronggang Yu, Peter Stone, "Performance analysis of a counter-intuitive automated stock-trading agent" Proceedings of the 5th international conference on Electronic Commerce (September 2003) p. 40.
http://portal.acm.org/ft_gateway.cfm?id=948011&type=pdf&coll=GUIDE&dl=GUIDE&CFID=34254586&CFTOKEN=56906445
[56] Id. at p. 41.

Chaturvedi and Mehta address this problem, noting that rational agents cannot always act as arbitrators for irrational agents. Rather they have to develop rational strategies for coping with the irrationality of other traders. [57] Again, they take an agent based approach.

I would like to suggest however that the problem of irrational stock market traders is no different from the problem of an irrational game player in a zero sum game. Suppose we are playing NIM (or tic tac toe known in England as noughts and crosses) and we use the minimax algorithm. But our opponent does not! Minimax is still the best algorithm for us to use. In a deterministic game it is clear that a player who does not employ minimax is likelier to lose and likelier to lose more quickly. Even in a game with random factors, the expectiminimax function is always superior to any other strategy over time. A player who does not employ minimax (or expectiminimax) will simply lose more often and more quickly than a player who does. That is, the player who employs an irrational strategy will merely lose more quickly.

In the stock market this principle translates as follows: they who trades based on irrational strategies will likely find themselves losing money. The irrational trader buys when all others are buying, and sells when all others are selling. They buy dear and sell cheaply. Thus they lose their investments. Had they been patient and held onto the stock they bought when it was overvalued and waited until the fundamental values of the company grew into their earlier expectations they would have been able to recover their paper losses and even (eventually...) make a profit. Unless the company fails or is beset with fraud, a buy and hold strategy is, eventually, profitably (though the eventuality may take years to materialize).

My hypothesis is that irrational actors tend to buy when the market is rising and sell when it is falling – even this can be profitable, so long as buying occurs at troughs and selling at peaks. If irrational actors do act „as a herd" buying and selling en masse i.e. as a group then that would lead to the conclusion that irrational agents behavior merely amplifies the magnitude of stock market booms and crashes. Just as the stock market in September 1929 was overvalued, so was it also in 1933 undervalued! It did take nearly twenty years for the stock market losses of 1929 to be regained – but a buy and hold strategy would have ultimately prevailed even against the worst stock market crash in history.

---

[57] Wei T. Yue, Alok R. Chaturvedi, Shailendra Mehta, "Is more information better? The effect of traders' irrational behavior on an artificial stock market", Proceedings of the twenty first international conference on Information systems (December 2000) p. 660. http://portal.acm.org/ft_gateway.cfm?id=359921&type=pdf&coll=GUIDE&dl=ACM&CFID=34254547&CFTOKEN=8281921

The good news for potential investors is: virtually every strategy is a winner. In the long term the stock market has consistently risen. The question is not whether you can make money by saving it – even a passbook savings account gives some return. Rather it is what strategies will lose you money (buying high, selling low, and not holding onto paper losses until they are erased). Regular (e.g. 100 euros per month), long term investing (during 20 or more years) will always yield a handsome return due to compounded interest.

An Agent Model for Stock Trading

The model I present uses six very simple agents. None of the agents „learn". However each agent has an individual strategy based on fundamental analysis, technical analysis, or both. The agents generally use fundamental analysis as that yields an objective measure of the worth of a given company. Technical analysis is used for two of the agents. The hypothesis is that the agents trading based on fundamental analysis will do better than those using technical analysis. The agents are:

Bears:
„Bears" are fundamentally sceptical. They require a favorable price-earnings ratio (p:e < 30), an undervalued stock (book value of the stock < market value of the stock), a low debt:equity ratio (<1) and positive earnings (a profitable company) before purchasing a stock.

Conservatives:
Conservative investors are less wary than bears but stilll conservative in their investment. Thus the stock mus have a low price:earnings ration (<30) and be undervalued (book value < market value).

Blue Chip Investors:

The blue chip agent is defined very simply in contrast. A blue chip investor is seeking return on dividends. They are not seeking to obtain a speculative gain on the fluctuation in price of the stock. Thus, essentially, the blue chip investor will look at the annual dividends of the company. In the agent model I used here, the blue chip investor only looks at the dividends of the company. If the dividends are high (i.e. over one dollar per share) the blue chip investor will buy the stock and then hold it to gain the income from the dividends. The blue chip agent will sell its stock immediately if the stock no longer yields dividends greater than one dollar per share.

Bargain Hunters:

The bargain hunter agent is also very simply defined. The bargain hunter looks to see whether the stocks value „on paper", i.e. the value assigned by the market, is less than the book value of the company. Book value is determined by comparing company assets to company debts. Thus a company with a market value less than its book value is undervalued. During economic booms it is difficult to find undervalued companies! Thus this strategy is really only effective during market troughs. Which means it is a good buying strategy when it can be used - but cannot always be used. If the book value of the share goes below the market value then the „bargain hunter" agent will immediately sell the stock.

Fools:

The fool agent will purchase any stock provided the price:earnings ratio is favorable, i.e. below 30. Similarly, if p:e goes above 30 the fool agent will sell. Clearly, some stock investors only look at p:e. However they also look at their purchase and sales price when selling. Thus this agent could be improved. For example:

 if p:e>30 and purchase<sales then sell

would be a better sales routine for the fool agent.

Idiots:

The idiot agent engages in no fundamental analysis. Instead it merely purchases and sells based on the market trends. The idiot agent presumes (wrongly!) that tomorrow will be like today. Thus if the market is going up the idiot agent will buy. If the market is going down the idiot agent will sell. It is perhaps frightening, but even the idiot agent will sometimes make good trades.

None of these agents represents the author's investment tactics. I would say that an eric agent would:

Only buy stocks with a p:e <30
Only buy stocks lower than or within 10% of their book value
Only buy stocks from companies making profits this year
Only buy stocks with a low debt-equity ratio (<1)
And sell any stock which has gained over 20% of its purchase price.

Of course, the eric agent will also miss trading opportunities! This is because it only buys companies which are undervalued or nearly so. Moreover, the eric agent misses profit opportunities. It is not at all unheard of for a stock to gain hundreds of percentage points in value! However, the eric agent also misses lots of opportunities for losses. First, it only buys stocks with solid fundamentals – a fact missed by the fools or idiots agents. It does not take as large a profit as it could, but it also avoids waiting too long to lock in profit opportunities.

The eric agent could be improved, for example, by factoring technical analysis in. The eric agent could expect: high oil prices and inflation to result in paper gainst in stock prices and real decline in productivity. Those, interestingly, are current factors in today's economy!

Future Research

The advantage of a neural network is that it is able to be trained. Its algorithms do not have to be hard coded. Further, the neural network can easily learn and adapt to new behaviors. However, a lot of the knowledge base in stock market investing can be hand coded.

Future research in agent based stock market trading should look at the following issues:
*further developing agent trading algorithms such as the reverse strategy, bears, blue chips and

convervatives to include:

a) technical analysis, i.e. how the market performance overall influences trading in a specific stock. Most existing research is only concentrating on technical analysis. Fundamentals analysis approaches should not ignore the points made by technical analysis. Fundamental and technical analysis need to complement each other.

b) learning procedures to allo the agents to i) learn about the market ii) learn about the other agents

c) opponent modeling in the stock market. For example, in bear markets the majority of agents may be acting like the bear agent I present. In bull markets the majority of agents may be acting like the fools agent I present. Opponent modeling could take this into account so that the eric agent knows that most other traders are now, say, conservative, and thus it will be a bad time to sell any security.

d) include a critic agent to evaluate the trading strategies of the various agents to try to develop a best trading strategy from the different strategies.